\documentclass[onecolumn,usenatbib]{mn2e}
\usepackage{epsfig}
\begin{document}
%%%%%%%%%%%%%%%%%%%%
%% Title page
%%%%%%%%%%%%%%%%%%%%
\title[Magnetosphere with Two-fluid Plasma Flows ]
{Numerical Construction
of Magnetosphere with Relativistic Two-fluid Plasma Flows }
\author[Y. Kojima and J. Oogi]{Yasufumi Kojima\thanks{%
E-mail:kojima@theo.phys.sci.hiroshima-u.ac.jp} 
and Junpei Oogi \\
Department of Physics, Hiroshima University, Higashi-Hiroshima, 
739-8526, Japan }
\maketitle

\begin{abstract}
%%%%%%%%%%%%%%%%%%%%%%%%%%%%%%%%%%%%%%%%%%%%%%%%%%%%%%%%
   We present a numerical model in which a cold pair plasma is ejected 
with relativistic speed through a polar cap region and flows almost radially
outside the light cylinder. Stationary axisymmetric structures of
electromagnetic fields and plasma flows are self-consistently calculated. In
our model, motions of positively and negatively charged particles are
assumed to be determined by electromagnetic forces and inertial terms, 
without pair creation and annihilation or radiation loss. 
The global electromagnetic fields are calculated by the Maxwell's equations
for the plasma density and velocity, without using ideal MHD condition.
Numerical result demonstrates the acceleration and deceleration of plasma
due to parallel component of the electric fields. 
Numerical model is successfully constructed for
weak magnetic fields or highly relativistic fluid velocity,
i.e, kinetic energy dominated outflow.
It is found that appropriate choices of boundary conditions and plasma
injection model at the polar cap should be explored in order
to extend present method  to more realistic pulsar magnetosphere,
in which the Poynting flux is dominated.
\end{abstract}
%%%%%%%%%%%%%%%%%%%%%%%%%%%%%%%%%%%%%%%%%%%%%%%%%%%


%(1)%%%%%%%%%%%%%%%%%%%%%%%%%%%%%%%%%%%%%%%%%%%
  \section{INTRODUCTION}
%%%%%%%%%%%%%%%%%%%%%%%%%%%%%%%%%%%%%%%%%%%%%%%
  A global structure of pulsar magnetosphere is one of key issues to
understanding energy outflow to the exterior. The numerical model has been
successfully developed in the past decade, although the basic equation was
already derived in the early days of pulsar theory. 
Extensive reviews are available in some books 
(e.g, \cite{Mic91,BGI93,Mes99}). 
\cite{CKF99} calculated the stationary axially symmetric magnetosphere
based on the force-free approximation. They for the first time showed
a solution with dipole magnetic field lines near a neutron star, 
which smoothly pass through the light cylinder to the wind region 
at infinity. In the model,
there is a current sheet flowing on the separatrix and equator outside 
the light cylinder. The magnetosphere model is subsequently explored
in detail by several authors; some physical properties represented
by the solution \citep{OK03}, the
Y-point singularity between open and closed field lines on 
the equator \citep{Uzd03,Tim06}, the electromagnetic luminosity 
in high numerical resolution \citep{Gru05,Gru06}.
Numerical construction of the magnetosphere around an
aligned rotator is also performed using time-dependent codes 
with the force-free and MHD approximations \citep{Kom06,McK06}. 
A stationary state, which is very similar to 
the solution given by \cite{CKF99}, is obtained with 
certain initial and boundary conditions.
The approach is extended to an oblique rotator by 
3D simulation codes \citep{Spi06,KalCon08}.

%%%
   An ideal MHD condition $\vec{E}= \vec{B} \times \vec{v}/c $ 
is used to determine the electric field in these calculations 
irrespective of the numerical methods.
Consequently, two electromagnetic field vectors
are always orthogonal $\vec{B} \cdot \vec{E} =0$,
and the parallel component of $\vec{E}$ 
along the plasma motion vanishes everywhere. 
This condition holds if the
plasma density much exceeds the Goldreich-Julian density \citep{GJ69}. 
The global structure based on the force-free and MHD approximations 
is obviously a good step to understanding the whole magnetosphere. 
However, it is important to study how and where the condition breaks down,
and how this changes the electromagnetic field structure and plasma
behavior. An alternative approach, in which the ideal MHD condition 
is relaxed, is necessary to address these problems. 
Breakdown of ideal MHD condition in the pulsar magnetosphere is 
qualitatively pointed out in the literature, e.g, \citep{MS94, GMMW04}.
It is our purpose to further study the problem by actual modeling.
It is necessary to determine the electric fields from
the distribution  of the charge density,
since $\vec{E}= \vec{B} \times \vec{v}/c $
is no longer used.
The electric acceleration or deceleration of fluids will be allowed
elsewhere,  since $ \vec{v} \cdot \vec{E} \ne 0$.
The location may be important to the observation.

%%%
  In this paper, we present an approach based on a two-fluid 
plasma consisting of positively and negatively charged particles. 
In this approach, the electromagnetic fields are modeled by Maxwell's 
equations with a plasma source. 
Resultant electric fields are not in general perpendicular to
magnetic fields $\vec{B} \cdot \vec{E} \ne 0$, i.e,
breakdown of ideal MHD condition.
The stream lines of the plasma flows are not a priori 
assumed to coincide with the magnetic field lines. The
plasma flows are determined by equation of motion under the electromagnetic
forces. In this paper, we consider a simple plasma model, 
a cold dissipationless plasma, 
in which thermal pressure, pair creation or annihilation and 
radiation loss are neglected.
If we further neglect the inertial terms in equations of motion,
then we have the force-free and MHD conditions \citep{GMMW04}. 
We here keep the inertial terms, in order to model the magnetosphere
taken into account one of non-ideal MHD effects.
We also assume stationary axisymmetric states 
in the electromagnetic fields and the plasma flows. However, the azimuthal
component of a vector is not zero in general. For example, a toroidal
magnetic field $B_\phi$ can be produced by poloidal currents 
$\vec{j}_p=(j_r, j_\theta )$, 
in the spherical coordinates $(r, \theta, \phi)$. 
This approach can be naturally extended to include other physical mechanisms
in the future. This paper is, therefore, the first step towards including
more physical processes.

%%%
  This paper is organized as follows. In section 2, we discuss our numerical
method to solve electromagnetic fields and fluid streams in two-dimensional
meridian plane. The relevant boundary conditions are also given. 
A model of plasma injection is given at the inner boundary,
a polar cap region.
The angular dependence of the injection rates is different between
positively and negatively charged particles, 
but the total is the same. The total current from
the polar cap is therefore zero.
This provides a model to construct magnetosphere with
charge-separated plasma flow. 
This choice of the injection model is not unique.
In section 3, we show our results of the global structure. 
Section 4 contains our conclusions.

%(2)%%%%%%%%%%%%%%%%%%%%%%%%%%%%%%%%%%%%%%%%%%%
  \section{ASSUMPTIONS AND EQUATIONS}
%%%%%%%%%%%%%%%%%%%%%%%%%%%%%%%%%%%%%%%%%%%%%%%
%(2.1)%%%%%%%%%%%%%%%%%%%%%%%%%%%%%%%%%%%%%%%%%
  \subsection{ Electromagnetic fields and plasma flows}
%%%%%%%%%%%%%%%%%%%%%%%%%%%%%%%%%%%%%%%%%%%%%%%
   Axially symmetric electromagnetic fields in the stationary state are
expressed by three functions 
$\Phi (r,\theta )$,$G(r,\theta )$,$S(r,\theta )$ as 
\begin{eqnarray}
\vec{E} &=&-\vec{\nabla}\Phi , \\
\vec{B} &=&\frac{1}{R}\vec{\nabla}G\times \vec{e}_{\phi }
+\frac{S}{R}\vec{e}_{\phi },
\end{eqnarray}
where we use spherical coordinate $(r,\theta ,\phi )$ and $R=r\sin \theta $.
It is convenient to use the non-corotational potential $\Psi =\Phi
-\Omega G/c$, where $\Omega $ is angular velocity of a central star.
Maxwell equations with charge density $\rho _{e}$, and poloidal and
toroidal components of current $(\vec{j}_{p},j_{\phi })$ are given by 
\begin{equation}
{\mathcal{D}}G=-\frac{4\pi }{c}Rj_{\phi },
  \label{eqn.Gfn}
\end{equation}
\begin{equation}
\frac{1}{R}\vec{\nabla}S\times \vec{e}_{\phi }=\frac{4\pi }{c}\vec{j}_{p},
  \label{Maxw.poloidal}
\end{equation}
\begin{equation}
\nabla ^{2}\Psi =-4\pi 
 \left( \rho _{e}-\frac{\Omega R}{c^{2}}j_{\phi}\right) 
-\frac{2\Omega }{cr^{2}}\left( r\frac{\partial G}{\partial r}+\cot
\theta \frac{\partial G}{\partial \theta }\right) , 
 \label{eqn.Psi}
\end{equation}
where operators ${\mathcal{D}}$ and $\nabla ^{2}$ in spherical coordinate
are given by 
\begin{equation}
{\mathcal{D}}=\frac{\partial ^{2}}{\partial r^{2}}
+\frac{\sin \theta }{r^{2}}\frac{\partial }{\partial \theta }
\left( \frac{1}{\sin \theta }\frac{\partial }{\partial \theta }\right) ,
\end{equation}
\begin{equation}
\nabla ^{2}=\frac{1}{r^{2}}\frac{\partial }{\partial r}
\left( r^{2}\frac{
\partial }{\partial r}\right) +\frac{1}{r^{2}\sin \theta }
\frac{\partial }{\partial \theta }
\left( \sin \theta \frac{\partial }{\partial \theta }
\right) .
\end{equation}

%%%
  We adopt a treatment in which the plasma is modeled as 
a two-component fluid. Each component, consisting of positively 
or negatively charged particles, is described by a number density
$n_{\pm }$ and velocity 
$\vec{v}_{\pm } = \vec{v}_{\pm p} + v_{\phi} \vec{e}_{\phi}$. 
Note that the proper density $n_{\pm }^{\ast }$ is related with
the lab-frame density $n_{\pm }$ by 
$n_{\pm }^{\ast }=n_{\pm }/\gamma _{\pm} $, 
where $\gamma _{\pm }$ is a Lorentz factor 
$\gamma _{\pm }=(1-(v_{\pm}/c)^{2})^{-1/2}$ (e.g, \cite{GMMW04}). 
We assume that the positive particle
has mass $m$ and charge $q$, while the negative one has mass $m$ and 
charge $-q$. 
The charge density and electric current are given in terms of 
$n_{\pm }$ and $\vec{v}_{\pm }$ as 
\begin{eqnarray}
\rho _{e} &=&q(n_{+}-n_{-}), 
   \label{dfn.charge} \\
\vec{j} &=&q(n_{+}\vec{v}_{+}-n_{-}\vec{v}_{-}).
   \label{dfn.current}
\end{eqnarray}
Continuity equation for each component in the stationary 
axisymmetric conditions is 
\begin{equation}
0=\vec{\nabla}\cdot (n_{\pm }\vec{v}_{\pm })
 =\vec{\nabla}\cdot (n_{\pm }\vec{v}_{\pm p}) .
\end{equation}
The poloidal velocity components $\vec{v}_{\pm p} $
are satisfied by introducing a stream function 
$F_{\pm}(r,\theta )$ as
\begin{equation}
n_{\pm }\vec{v}_{\pm p}=\frac{1}{R}\vec{\nabla}F_{\pm }\times \vec{e}_{\phi }.
  \label{dfn.fluid}
\end{equation}
From the definition, the number density is given by 
\begin{equation}
n_{\pm }=\frac{|\nabla F_{\pm }|}
              {R(v_{\pm r}^{2}+v_{\pm \theta }^{2})^{1/2}} .
  \label{dfn.density}
\end{equation}
From eqs.(\ref{dfn.current}) and (\ref{dfn.fluid}), 
the current function $S$
in eq.(\ref{Maxw.poloidal}) can be solved as 
\begin{equation}
S=\frac{4\pi q}{c}(F_{+}-F_{-}).
\end{equation}

%%%
   The electromagnetic force is dominant so that collision, 
thermal pressure and gravity are ignored. The interaction between 
two-component fluids is assumed only through the global 
electromagnetic fields. The equation of motion for each component 
with mass $m$ and charge $\pm q$ in the stationary state is given by 
\begin{equation}
\left( \vec{v}_{\pm }\cdot \vec{\nabla}\right) \gamma _{\pm }\vec{v}_{\pm
}=\pm \frac{q}{m}
\left[ \vec{E}+\frac{\vec{v}_{\pm }}{c}\times \vec{B}\right] .
  \label{motion.eqn}
\end{equation}
By adding and subtracting equations (\ref{motion.eqn}) for two components, 
we have an equation of one-fluid bulk motion and 
a generalized Ohm's law. See e.g, \citep{MM96,GMMW04} 
for the detailed discussion.  We do not follow such a treatment, 
but rather solve eq.(\ref{motion.eqn}) for
each component. Using the identity 
$(\vec{v}\cdot \vec{\nabla})\gamma \vec{v}$ 
$=(\vec{\nabla}\times \gamma \vec{v})\times
\vec{v}+\vec{\nabla}\gamma c^{2}$,
we find two conserved quantities along each stream line,
corresponding to axially symmetric and stationary conditions.
They are generalized angular momentum $J_{\pm }$
and Bernoulli integral $K_{\pm }$, 
which are obtained by
the azimuthal component of eq.(\ref{motion.eqn})
and a scalar product of $\vec{v} $ and eq.(\ref{motion.eqn})
\citep{Mes99}.
Their explicit expressions are given by
\begin{equation}
J_{\pm }=\gamma _{\pm }v_{\pm \phi }R\pm \frac{q}{mc}G ,
   \label{const.Ang}
\end{equation}
\begin{equation}
K_{\pm }=\gamma _{\pm }\pm \frac{q}{mc^{2}}\Phi .
   \label{const.Bern}
\end{equation}
These quantities depend on the stream functions $F_{\pm }$ only, 
and the spatial distributions are therefore determined by $F_{\pm }$ 
which is specified at the injection point in our model.
The convenient form for the third component of eq.(\ref{motion.eqn})
is the azimuthal component of a cross product of
$\vec{v} $ and eq.(\ref{motion.eqn}). This means
a perpendicular component to the stream lines, which is  given by
\begin{equation}
{\mathcal{D}}F_{\pm }=\vec{\nabla}\ln \left( 
   \frac{n_{\pm }}{\gamma _{\pm }}\right)
\cdot \vec{\nabla}F_{\pm }
+\frac{c^{2}n_{\pm }^{2}R^{2}}{\gamma_{\pm }}
\left( K_{\pm }^{\prime }
 -\frac{v_{\pm \phi }}{c^{2}R}J_{\pm}^{\prime }\right)
 \pm \frac{q}{mc}\frac{n_{\pm }}{\gamma _{\pm }}S,
  \label{eqn.trans1}
\end{equation}
where $J_{\pm }^{\prime }$ and $K_{\pm }^{\prime }$ are derivatives of 
$J_{\pm }$ and $K_{\pm }$ with respect to $F_{\pm }$. Using 
$\vec{\nabla}J_{\pm}=J_{\pm }^{\prime }\vec{\nabla}F_{\pm }$,
$\vec{\nabla}K_{\pm }=K_{\pm }^{\prime }\vec{\nabla}F_{\pm }$ 
and eq.(\ref{dfn.density}), eq.(\ref{eqn.trans1}) can be written in
an alternative form 
\begin{equation}
{\mathcal{D}}F_{\pm }=\left[ 
\vec{\nabla}\ln \left( \frac{n_{\pm }}{\gamma_{\pm }}\right) 
+\frac{c^{2}}{\gamma _{\pm }(v_{\pm r}^{2}
+v_{\pm \theta}^{2})}
\left( \vec{\nabla}K_{\pm }-\frac{v_{\pm \phi }}{c^{2}R}\vec{\nabla}
J_{\pm }\right) \right] \cdot 
\vec{\nabla}F_{\pm }\pm \frac{q}{mc}\frac{n_{\pm }}{\gamma _{\pm }}S.
   \label{eqn.trans2}
\end{equation}

%%%
  The stream function can not be defined in corotating region, 
where poloidal components of the velocity vanish, and the expression
(\ref{dfn.fluid}) is no longer used. Instead, 
the charge density $\rho _{e\ast }$ and current $\vec{j}_{\ast }$ 
are given in terms of the corotating condition 
$\Phi =\Omega G/c$, 
$\vec{j}_{\ast } =\rho _{e\ast }\Omega R\vec{e}_{\phi }$. 
From eq.(\ref{eqn.Psi}), the
corotating charge density is given by 
\begin{equation}
4\pi \rho _{e\ast }=\frac{2c\Omega }{(c^{2}-\Omega ^{2}R^{2})R}\left( \sin
\theta \frac{\partial G}{\partial r}
+\frac{\cos \theta }{r}\frac{\partial G}{\partial \theta }\right) .
\end{equation}

%%%
 Three velocity components,  Lorentz factor and number density
are determined by eq.(\ref{dfn.density}) and two integrals 
(\ref{const.Ang}),(\ref{const.Bern}), if four functions 
$G$, $\Psi $ and $F_{\pm }$ are known.
The charge density (\ref{dfn.charge}) and current (\ref{dfn.current}) 
are calculated from these fluid quantities of both species.
In the corotating region, they are given by corotating charge density
and current.
Irrespective of the spatial region,  
the source terms of partial 
differential equations for $G$, $\Psi $ and $F_{\pm }$
depend on themselves in a non-linear manner. 
Some iterative methods are needed to self-consistently solve a set of 
eqs.(\ref{eqn.Gfn}),(\ref{eqn.Psi}) and (\ref{eqn.trans2}). There is no
established method so far to solve nonlinearly coupled equations, 
so that our numerical procedure is rather primitive.
Initial guess for these functions is
assumed, say $G ^{(0)}$, $\Psi ^{(0)}$ and $F_{\pm }^{(0)}$. Using these
functions, the source terms are calculated, and a new set of functions 
$G^{(1)}$, $\Psi ^{(1)}$ and $F_{\pm }^{(1)}$ are solved from these source
terms with appropriate boundary conditions. The procedure is repeated until
the convergence, say, $|G^{(n+1)}-G^{(n)}|$,
$|\Psi ^{(n+1)}-\Psi^{(n)}|$,
$|F_{\pm}^{(n+1)}-F_{\pm }^{(n)}|<\varepsilon $, 
where $\varepsilon $ is a small number. The iteration scheme may not 
necessarily lead to a convergent
solution, since there is no mathematical proof.

In order to examine our numerical scheme, we have performed 
a test for the split-monopole case, 
for which an analytic solution is known \citep{Mich73}.
The non-corotational electric potential in the solution is zero 
everywhere, so that the condition $\Psi =0$ is used and a reduced 
system of $G$ and $ F _{\pm}$  is checked.
These functions are numerically solved by a finite difference method
with appropriate boundary conditions in the upper half plane.
Results for the  convergence to the solution are given in Table \ref{table1}.
Two types of initial trial functions and two different grid numbers
are used.
Deviation from the analytic solution
is shown by a norm  $\Vert \delta f \Vert _{n}$,
which is  evaluated at all grid points as
$ \Vert \delta f \Vert _{n} 
= \left[ \sum ( f ^{(n)} (r_i, \theta_j) -f^{\ast}(r_i, \theta_j) )^2  
\right] ^{1/2} 
/\left[ \sum ( f^{\ast} (r_i, \theta_j) )^2  \right] ^{1/2} $,
where $ f^{\ast}  $ is the analytic solution and $ f ^{(n)} $ 
is  numerical result after $n$ iterations.
We have repeated until the relative error 
$\varepsilon=1 \times 10^{-3} $ in this test problem.
We have started from $  G^{(0)} =0 $, so that
$\Vert \delta G \Vert _{0} =1$.
The initial choice of $G^{(0)}$ is not so important, since
the numerical solution approaches the analytic one 
at the first step.
On the other hand, the choice of 
initial guess for  $ F _{\pm}$ is important.
It is not easy to set large deviation at the initial step, 
since the function  $ F _{\pm}$ should be monotonic.
If there is a maximum or minimum, where $ |\nabla F _{\pm} | =0$
inside the numerical domain, the flow vanishes $ n v_{p} =0$.
This causes a numerical difficulty at that point.
From the monotonic nature consistent with the
boundary conditions,  the initial norm
$\Vert \delta  F _{\pm} \Vert _{0}$ can  not be large.
Table \ref{table1} shows that the numerical solutions
successfully converge on the analytic ones within certain  errors.
Convergence factor $\varepsilon $ does not exactly correspond 
to deviation from true solution, but gives an estimate.
The true solution is not known in most problems, 
and the deviation can not be calculated.
The convergence factor $\varepsilon $ can be regarded as
error estimate.

%%%%%%%%%%%%%%%%%%%%%%%%%%%%%%%%%%%%%%%%%%%%%%%%%%%%%%%%%%%%%%%%%%%%%%
% Table 1
%%%%%%%%%%%%%%%%%%%%%%%%%%%%%%%%%%%%%%%%%%%%%%%%%%%%%%%%%%%%%%%%%%%%%%
\begin{table}
\caption{ Convergence test }
\begin{center}
\begin{tabular}{cccccccc}
\hline\hline
Model & grid & 
 $\Vert \delta G \Vert _{0}$ &
   $\Vert \delta F_{+} \Vert_{0} $ & $\Vert \delta F_{-} \Vert_{0} $ &
     $\Vert \delta G \Vert _{N}$  &
      $\Vert \delta F_{+} \Vert _{N}$ & $\Vert \delta F_{-} \Vert _{N}$ \\
\hline
A$_1$ & 150 $\times$ 50 &
 1.0 &  0.15  & 0.15 &
 1.4 $\times$ 10$^{-3}$ &
 1.3 $\times$ 10$^{-2}$ & 1.1 $\times$ 10$^{-2}$  \\
A$_2$ & 300 $\times$ 100 &
 1.0 &  0.15  & 0.15 &
 4.1 $\times$ 10$^{-4}$ &
 7.4 $\times$ 10$^{-3}$ & 5.6 $\times$ 10$^{-3}$  \\
B$_1$ & 150 $\times$ 50 & 
 1.0 &  0.25  & 0.25 &
 2.5 $\times$ 10$^{-3}$ &
 1.6 $\times$ 10$^{-2}$ & 1.2 $\times$ 10$^{-2}$  \\
B$_2$ & 300 $\times$ 100 & 
 1.0 &  0.25  & 0.25 &
 9.1 $\times$ 10$^{-4}$ &
 7.8 $\times$ 10$^{-3}$ & 5.2 $\times$ 10$^{-3}$  \\
\hline\hline
\end{tabular}
\end{center}
\label{table1}
\end{table}
%%%%%%%%%%%%%%%%%%%%%%%%%%%%%%%%%%%%%%%%%%%%%%%%%%%%%%%%%%%%%%%%%%%%%%

%(2.2)%%%%%%%%%%%%%%%%%%%%%%%%%%%%%%%%%%%%%%%%%
   \subsection{ Boundary conditions}
%%%%%%%%%%%%%%%%%%%%%%%%%%%%%%%%%%%%%%%%%%%%%%%
  We assume that the axially symmetry around a polar axis and 
the reflection symmetry across an equator. The numerical domain in 
the spherical coordinate $(r,\theta )$ is 
$r_{0}\leq r\leq r_{1}$, $0\leq \theta \leq \pi /2$. 
The inner and outer radii in our calculations are $r_{0}=r_{L}/5$ and 
$r_{1}=5r_{L}$, where $r_{L}=c/\Omega $ is the distance to the light
cylinder. Figure 1 schematically represents the numerical domain. 
The functions at the
inner boundary $r_{0}$ are closely related with plasma injection model,
which is separately discussed in the next subsection. 
We here discuss the boundary
conditions at the axis, equator and outer radius.

%%%
   We solve the magnetic flux function $G$ in the upper half plane 
between $r_{0}$ and $r_{1}$, which is the region enclosed by 
a curve $PBRQP$ in Fig.1. Poloidal magnetic field at the inner 
boundary $r_{0}$, i.e, on $PB$
is dipole, so that we impose the condition, 
$G=\mu \sin ^{2}\theta /r_{0}$, where $\mu $ is the
magnetic dipole moment. The plasma is injected through a polar 
cap region $0\leq \theta \leq \theta _{0}$ at $r_{0}$, $PA$ in Fig.1. 
A curve $AL$ represents the last open magnetic field line. 
All the field lines originated
from a point with $\theta \leq \theta _{0}$ at $r_{0}$ extend to infinity,
whereas those from $\theta >\theta _{0}$ are closed. 
The point $L$ is $(r_{L},\pi /2)$.
The last open line is given by 
$G_{0}=\mu \sin ^{2}\theta_{0} /r_{0}$. 
For purely dipolar field, the polar cap region and 
critical field line are given by 
$\sin ^{2}\theta_{0} = r_{0}/r_{L}$ 
and $G_{0} =\mu /r_{L}$. These values in our numerical model
are not known a priori, but are determined simultaneously with 
the global structures.
The boundary condition on the equator $\theta =\pi /2$ is 
$\partial G/\partial \theta =0$ inside the light cylinder,
i.e, on $BL$ in Fig.1,
while outside it $G=G_{0}$.
This condition on the equator means $B_r =0 $
inside the light cylinder, but $B_\theta =0 $ outside it. 
The boundary condition at the outer radius on $QR$ is continuous, 
$\partial G/\partial r=0$
at $r_{1}$. This condition means that the poloidal magnetic field becomes
radial, since $B_\theta =0 $. On the polar axis $PQ$, 
we impose the regularity condition which is given by 
$G\propto (r\sin \theta )^{2}$ for 
$\theta \rightarrow 0$, i.e, $B_\theta =0 $ on the axis.

%%%
   Next, we consider the boundary conditions for the stream 
function $F_{\pm }$,
which is defined outside of the corotation region, i.e,
a region enclosed by a curve $PALRQP$ in Fig.1. We here assume that 
irrespective of the fluid species, the last stream line 
coincides with the last open line of the poloidal magnetic field. 
The function $F_{\pm }$ should continuously approach a constant 
$F_{\pm c}$ on the boundary with the corotation region $AL$. 
Outside the light cylinder, the boundary condition on the equator $LR$ 
is $F_{\pm }=F_{\pm c}$ for $r_{L}\leq r\leq r_{1}$ 
at $\theta =\pi /2$, since there is
no flow across the equator due to the refection symmetry. 
Outer boundary condition at $r_{1}$ is 
$\partial F_{\pm }/\partial r=0$. This 
condition also means that the flow becomes radial
since $nv_{\theta }=(\partial F/\partial r)/R=0$. 
The regularity condition
on the polar axis $PQ$ is $F_{\pm }\propto (r\sin \theta )^{2}$ 
for $\theta\rightarrow 0$.

%%%
  Finally, we consider the boundary conditions for $\Psi $, 
which is non-corotating part of the electric potential. 
We solve it only outside the corotating region, the region
enclosed by a curve $PALRQP$ in Fig.1, since $\Psi =0$ in the corotating
region. As the boundary condition of $\Psi $, the function continuously
becomes zero, $\Psi =0$, toward the boundary $AL$ with the corotation region.
Outside the light cylinder, the condition on the equator $LR$ is 
assumed as $ \Psi =0$ at $\theta =\pi /2$ 
for $r_{L}\leq r\leq r_{1}$. 
Outer boundary condition at $r_{1}$ is also assumed as $\Psi =0$. 
One might think the boundary condition on the
polar axis $PQ$ is $\Psi =0$, for which ideal MHD condition
$\vec{B} \cdot \vec{E}=0$ is satisfied due to $E_r =B_\theta =0 $.
We found that the Dirichlet condition is too 
severe to lead to any numerical solution.
We use the regularity condition of $\Psi $,
that is, $\partial\Psi /\partial \theta =0$.
The Neumann condition 
is less severe, and allows the numerical solution.
This means $E_\theta =0 $ on the axis.
The ideal MHD condition is likely to be broken near the axis,
$\vec{B} \cdot \vec{E} \ne 0$.
This condition is quite different from usual MHD treatment.

%%%%% FIG1 %%%
\begin{figure}
\includegraphics[scale=0.5]{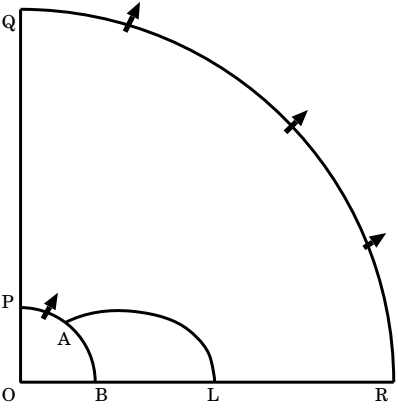}
\caption{ Meridian region of numerical calculation. The region enclosed by a
curve $ABLA$ is a corotating region. Plasma is injected through the polar
region $PA$ and goes out through the outer radius $QR$. A line $OQ$ is polar
axis, and $OR$ is an equator. 
\label{fig1}
}
\end{figure}
%%%%%%%%%%%%%%

%(2.3)%%%%%%%%%%%%%%%%%%%%%%%%%%%%%%%%%%%%%%%%%
  \subsection{ Injection model}
%%%%%%%%%%%%%%%%%%%%%%%%%%%%%%%%%%%%%%%%%%%%%%%
  In our model, plasma is assumed to flow through a small polar region, 
$0 \leq \theta \leq \theta _{0}$, at the inner boundary $r_{0}$, 
represented by $PA$ in Fig.1. 
If the stream lines completely agree with the magnetic dipolar field lines,
then the stream function is given by $F_{\pm} \propto \sin ^{2}\theta /r$
near the surface $r_{0}$. 
We assume that the function for each particle type slightly deviates 
from the dipolar configuration near the polar cap region. 
In our injection model, the stream function
in the range of $0\leq \theta \leq \theta _{0}$ 
is given by 
\begin{equation}
F_{\pm }(r_{0},\theta )=\frac{\lambda \sin ^{2}\theta }{r_{0}}
\left[ 1\mp
\alpha \left( 1-\left( \frac{\sin \theta }{\sin \theta _{0}}\right)
^{2}\right) \right] ,
    \label{eqn.inflow}
\end{equation}
and the current function $S$ is calculated as
\begin{equation}
S=-\frac{8\pi q\lambda \alpha }{cr_{0}}\sin ^{2}\theta 
\left[ 1-\left( \frac{\sin \theta }{\sin \theta _{0}}\right) ^{2}\right] .
    \label{eqnS.bc1}
\end{equation}
The constant $\alpha $ determines the deviation from the dipolar field. 
The poloidal current completely vanishes in the limit of $\alpha =0$, 
where the positively charged particles and negatively charged particles 
move along the common stream lines. 
The scale factor $\lambda $ is chosen as 
$4\pi q\lambda \alpha =\mu \Omega $, 
so that the current function can be written as
$S\approx -2\Omega G_{d}/c$ near the polar region, 
where $G_{d}=\mu \sin ^{2}\theta /r$. 
The current function corresponds to the split-monopole solution near 
the polar region \citep{Mich73}. 
Our current function $S$ smoothly goes to zero at the edge 
of polar cap, $ \theta =\theta _0 $.
This property is different from that of
the force-free and ideal MHD approximations
in which the function has a discontinuity \citep{CKF99,OK03,Gru05}. 
That is, there is a current sheet.

%%%
   The current density as a function of angle is calculated as 
$j_{r}=-4\mu \Omega
\cos \theta \lbrack 1-2(\sin \theta /\sin \theta _{0})^{2}]/r_{0}^{3}$
at the polar cap.
The electric current is negative for 
$0\leq \sin \theta <\sin \theta _{0}/\sqrt{2}$
and positive for 
$\sin \theta _{0}/\sqrt{2}<\sin \theta \leq \sin \theta_{0} $. 
The total current ejected through the polar cap region is zero, 
since $S(r_{0},\theta _{0})=0$. 
The positive or negative current flow is produced from
the charge-separated plasma, i.e, different number-density 
distribution between two components in our model.
The injection flow at $r_{0}$ is calculated from eq.(\ref{eqn.inflow}) as 
\begin{equation}
n_{\pm }v_{\pm r}=\frac{2\lambda \cos \theta }{r_{0}^{3}}\left[ 1\mp
\alpha \left( 1-2\left( \frac{\sin \theta }{\sin \theta _{0}}\right)
^{2}\right) \right] .
\end{equation}
We assume that the flow speed  $v_{0}$ is 
relativistic, $\gamma_{0}=(1-(v_{0}/c)^{2})^{-1/2}\gg 1$, 
and is independent of $\theta $. 
This property is assumed to be the same for each particle type. 
The flow direction through the small polar region $\cos \theta \sim 1$ 
is almost radial and the velocity is $v_{r}\sim c$, so that
the number density at $r_{0}$ is approximately given by 
\begin{equation}
n_{\pm }=\frac{2\lambda }{cr_{0}^{3}}\left[ 1\mp \alpha \left( 1-2\left( 
\frac{\sin \theta }{\sin \theta _{0}}\right) ^{2}\right) \right] .
  \label{eqndn.bc}
\end{equation}
The charge density is given by 
\begin{equation}
\rho _{e}=-\frac{4q\lambda \alpha }{cr_{0}^{3}}\left[ 1-2\left( \frac{\sin
\theta }{\sin \theta _{0}}\right) ^{2}\right] =-\frac{\mu \Omega }{\pi
cr_{0}^{3}}\left[ 1-2\left( \frac{\sin \theta }{\sin \theta _{0}}\right) ^{2}
\right] ,
  \label{eqnrho.bc}
\end{equation}
where our choice of parameter $4\pi q\lambda \alpha =\mu \Omega $ is used.
The typical value of eq.(\ref{eqnrho.bc}) 
$\mu \Omega /(\pi cr_{0}^{3})=B_{d}\Omega /(2\pi c)$ 
is the Goldrich-Julian charge density
for the field strength $B_{d}$ of magnetic dipole.

%%%
  Near the polar cap region, the force-free condition is satisfied, so that
the current function $S$ and electric potential $\Psi $ depend on the
magnetic flux function $G$.
We adopt the following forms, $S_{p}(G_{d})$ and $\Phi _{p}(G_{d})$, as a
function of dipolar flux function $G_{d}$ as 
\begin{equation}
S_{p}=-\frac{2\Omega G_{d}}{c}(1-bG_{d}),  \label{eqnS.bcp}
\end{equation}
\begin{equation}
\Phi _{p}=\frac{\Omega G_{d}}{c}+\frac{\Omega }{2bc}\left[ 1-(bG_{d})^{2}%
\right] ,  \label{eqnPh.bcp}
\end{equation}
where $b=r_{0}/(\mu \sin ^{2}\theta _{0})$. 
Equation (\ref{eqnS.bcp}) is reduced to
eq.(\ref{eqnS.bc1}) at $r_{0}$, and the Poisson
equation is approximately satisfied for
$\Phi _{p}$ (\ref{eqnPh.bcp}) and the charge density (\ref{eqnrho.bc}).
 The electric current in the force-free condition is generally given by 
\begin{equation}
\vec{j}=c\frac{dS}{dG}\vec{B}
+\rho _{e}cR\frac{d\Phi }{dG}\vec{e}_{\phi }.
  \label{eqnFF}
\end{equation}
By the straightforward calculations, it is found that
the poloidal components of eq.(\ref{eqnFF})
with the expressions $S_{p}(G_{d})$ and $\Phi _{p}(G_{d})$ 
are satisfied and toroidal component gives a small value 
$j_{\phi }\approx \rho _{e}c\times (r_{0}/r_{L})^{5/2}\ll j_{r}\approx
\rho _{e}c$. We regard $j_{\phi }=0$ and impose the $\phi $-component of
the fluid velocity as 
\begin{equation}
v_{+ \phi }=\frac{n_{-}+n_{+}}{2n_{+}}R_{0}\Omega ,
~~~~v_{- \phi }=\frac{n_{-}+n_{+}}{2n_{-}}R_{0}\Omega ,
  \label{eqnv3.bc}
\end{equation}
where $R_{0} =r_{0} \sin \theta$.
For this choice, total angular momentum of plasma flow is 
$m(n_{-}+n_{+})\gamma _{0}R_{0} ^{2}\Omega $
at the inner boundary.

%%%
  We here summarize the boundary conditions at $r_{0}$.
Equation (\ref{eqn.inflow}) is used for
$F_{\pm }$ with $4\pi q\lambda \alpha =\mu \Omega $, 
and eq.(\ref{eqnPh.bcp}) for $\Phi $.
Two integrals  $J_{\pm}$ and $K_{\pm}$ are calculated at $r_{0}$
as a function of polar angle $\theta$ 
from eqs.(\ref{eqndn.bc}),(\ref{eqnPh.bcp}),(\ref{eqnv3.bc}),
$G_d$ and $\gamma_0$. 
The relations $J_{\pm}(F_{\pm })$ and $K_{\pm}(F_{\pm })$ are constructed
by eliminating $\theta$ in terms of eq.(\ref{eqn.inflow}).

%(3)%%%%%%%%%%%%%%%%%%%%%%%%%%%%%%%%%%%%%%%%%%%
  \section{NUMERICAL RESULTS}
%%%%%%%%%%%%%%%%%%%%%%%%%%%%%%%%%%%%%%%%%%%%%%%
  We use a finite difference method to solve a set of partial differential
equations. The typical grid number in the spherical coordinate $(r,\theta )$
is $300\times 100$ for $0.2\leq r/r_{L}\leq 5$ 
and $0\leq \theta \leq \pi /2$. 
The polar cap region at the inner boundary is covered by approximately 
30 grid points. We have obtained the same result by changing the grid numbers
as $150\times 50$ or $450\times 150$. The convergent factor is 
$\varepsilon \approx 10^{-2}$, and can not be improved so much by 
the grid refinement. We demonstrate numerically 
constructed magnetosphere. 
Parameters used in the numerical calculation are  
a deviation parameter $\alpha =0.2$, Lorentz factor
$\gamma _{0}=10^{2}$ and $q\mu /(mc^{2}r_{L}^{2})=10$. 
The last dimensionless parameter is 
magnetic gyration frequency to angular velocity 
$\Omega =c/r_{L} $ of a star.
Numerical results of the plasma flows and
electromagnetic fields depend on a single combination, 
$\eta =q\mu /(mc^{2}\gamma _{0} r_{L}^{2})$, 
as far as $\gamma _{0}\gg 1$, since
the source terms for eqs.(\ref{eqn.Gfn}),(\ref{eqn.Psi}) 
and (\ref{eqn.trans2}) are scaled by it.
Thus $\eta $ is a key parameter to determining the global structure. 
See Appendix for the details.

%%%
  Figure \ref{fig2} shows numerical solution of the magnetic function $G$. 
We also show that of dipole field $G_{d}=\mu \sin ^{2}\theta /r$ 
for the comparison. The poloidal magnetic field is dipole near 
the inner radius $r_{0}/r_{L}=0.2$. The field gradually deviates 
from the dipole, and becomes open outside the
light cylinder. The field configuration is eventually radial near outer
radius $r_{1}/r_{L}=5$. 
The numerical result provides the last open field
line as $G_{0} \approx 1.15 \mu /r_{L}$. 
The critical value is $G_{0}=1.592 \mu /r_{L}$ in magnetosphere
fulled with rigidly rotating plasma \citep{Mich73,Mic91,MP92},
and $G_{0}=1.36\mu /r_{L}$ \citep{CKF99}, 
$G_{0}=1.27\mu /r_{L}$ \citep{Gru05} in a solution with the
force-free approximation, and 
$G_{0}=1.26\mu /r_{L}$ \citep{Kom06} 
in MHD simulation. 
Our value is smaller than that of other models, but 
is not so different.
The total current is different from 
that of these models, 
so that there is no reason why the critical value should agree.

%%%%% FIG2 %%%%%%%%%%%%%%%%%%%%%%%%%%%%%%%%%%%%
\begin{figure}
%%%%%%%%%%%%%%%%%%
\begin{minipage}{0.45\linewidth}
\includegraphics[scale=1.0]{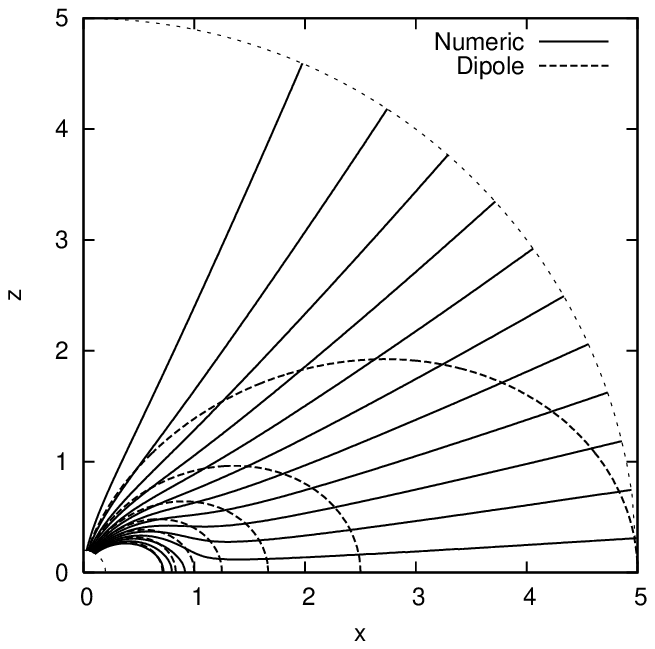}
\caption{  Magnetic flux functions.
Solid curves denote the flux surfaces 
of numerical solution for $ G r_{L}/\mu =$
$ 0.1, 0.2,\cdots, 1.4$, and
dotted curves those of dipole for $ G_{d} r_{L}/\mu =$
$ 0.2, 0.4,\cdots, 1.4$ starting from the polar axis.
\label{fig2}
}
\end{minipage}
\hspace{8mm} 
%%%%% FIG3 %%%%%%%%%%%%%%%%%%%%%%%%%%%%%%%%%%%%
\begin{minipage}{0.45\linewidth}
\includegraphics[scale=1.0]{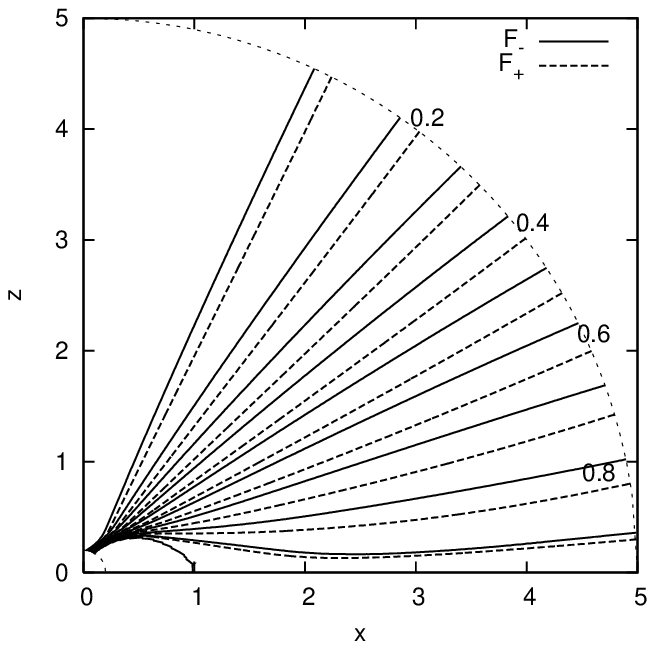}
\caption{ Stream functions for
negatively and positively charged fluids.
Solid curves denote the flux surfaces of  $ F_{-} $, and
dotted curves those of $ F_{+}$
in intervals of $ 0.1 (\lambda \sin^2 \theta_{0} /r_{0}) $.
\label{fig3}
}
\end{minipage}
\end{figure}
%%%%%%%%%%%%%%%%%%%%%%%%%%%%%%%%%%%%%%%%%%%%%%%

%%%
   Figure \ref{fig3} shows the results of the stream functions 
for both species. The global structure of stream lines is almost 
the same as that of the magnetic field lines shown in Fig.\ref{fig2}, 
although the numerical agreement is not so complete. 
Thus, flows in meridian plane is almost parallel to the magnetic
field lines. 
A difference between $F_{+}$ and $F_{-}$ at large radius 
originates from the inner boundary condition at $r_{0}$, 
where the fraction of negatively charged plasma is slightly larger 
in polar region $\theta \approx 0$, 
but smaller for $\theta \approx \theta_{0}$. 
The property extends to the outer radius. At large radius, 
the flow becomes is radial and the velocity is still relativistic, 
so that the number density decreases with the radius,
$n_{\pm} \approx |\partial _{\theta}
F_{\pm}|/(r^2 \sin \theta )$ $\propto r^{-2}$. 
However, the fraction $(n_{+} -n_{-})/(n_{+} +n_{-})$ is still
finite, and the charge separation remains.
Our numerical model shows that negatively and positively
charged regions are separated approximately by a curve with  
$F_{\pm} \approx 0.5 (\lambda \sin^2 \theta_{0} /r_{0}) $.

\begin{figure}
%%%%%%%%%%%%%%%%%%
\begin{minipage}{0.45\linewidth}
\includegraphics[scale=1.0]{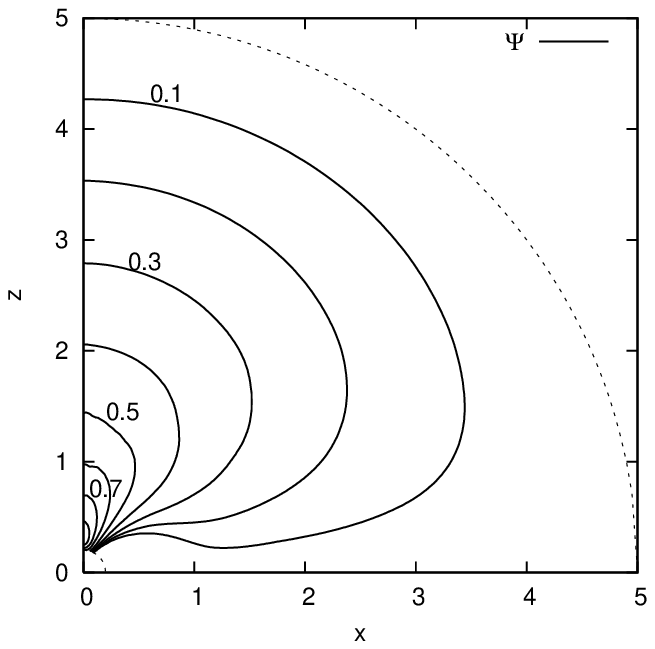}
\caption{  Contour of non-corotating part of
electric potential $\Psi=\Phi-\Omega G/c$. 
Contour levels outwardly decrease in intervals of
0.1$\mu \Omega /(c r_{L})$.
\label{fig4}
}
\end{minipage}
\hspace{8mm} 
%%%%% FIG5 %%%%%%%%%%%%%%%%%%%%%%%%%%%%%%%%%%%%
\begin{minipage}{0.45\linewidth}
\includegraphics[scale=1.0]{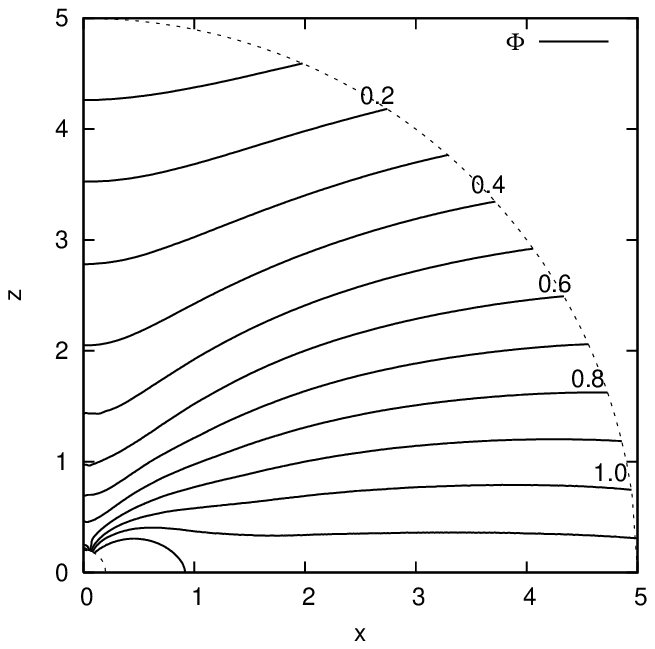}
\caption{  Contour of electric potential $\Phi$.
Contour levels upwardly decrease in intervals of
0.1$\mu \Omega /(c r_{L})$.
\label{fig5}
}
\end{minipage}
\end{figure}

%%%
   We show numerical result of the electric potential. 
The contour of the non-corotating part 
$\Psi = \Phi - \Omega G/c$ is shown in Fig.~\ref{fig4}.
There is a peak on the polar axis at $r \approx 0.4 r_{L}$. 
The function $\Psi $ decreases
toward the outer boundaries, where $\Psi=0$ is imposed 
as the boundary condition at outer radius, 
and on the last open magnetic field. 
Numerical result shows the maximum value 
$\Psi = 0.85 \mu \Omega/(c r_{L})$
at $(r,\theta)= ( 0.4 r_{L} , 0)$. This value is not small, 
since the maximum of corotating electric potential is 
$\Omega G/c  \approx 1.15 \mu \Omega/(c r_{L})$. 
Total electric potential $\Phi = \Psi + \Omega G/c$ is shown 
in Fig.~\ref{fig5}. 
Overall structure is very different from the magnetic flux 
function $G$  or stream functions $F_{\pm}$ shown 
in Figs.\ref{fig2}-\ref{fig3}. 
The difference is clear at the polar region, whereas the agreement 
becomes better at high latitude region near the equator.
Ideal MHD condition $\vec{B} \cdot \vec{E} = 0 $ is not assumed 
in our model. The deviation is very large on the polar axis.
This feature is closely related with the boundary conditions. 
As discussed in section 2, 
the boundary condition of $\Psi$
is not $E_{r}=0$, but $E_{\theta}=0$ on the axis. 
This mathematical condition may allow the
large value of $\Psi$ on the axis. On the other hand, 
$\Psi $ almost remains 
zero near the equatorial region from the boundary condition.

%%%%% FIG4 %%%%%%%%%%%%%%%%%%%%%%%%%%%%%%%%%%%%
%%%%%%%%%%%%%%%%%%%%%%%%%%%%%%%%%%%%%%%%%%%%%%%

\begin{figure}
%%%%%%%%%%%%%%%%%%
\begin{minipage}{0.45\linewidth}
\includegraphics[scale=1.0]{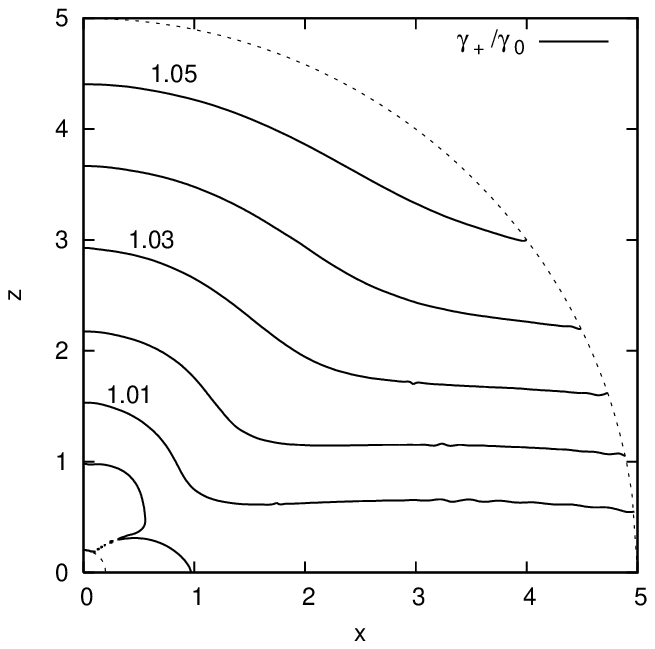}
\caption{  Increase of Lorentz factor
for positively charged particles.
Contour levels outwardly increase in intervals of
0.01. 
\label{fig6}
}
\end{minipage}
\hspace{8mm} 
%%%%% FIG7 %%%%%%%%%%%%%%%%%%%%%%%%%%%%%%%%%%%%
\begin{minipage}{0.45\linewidth}
\includegraphics[scale=1.0]{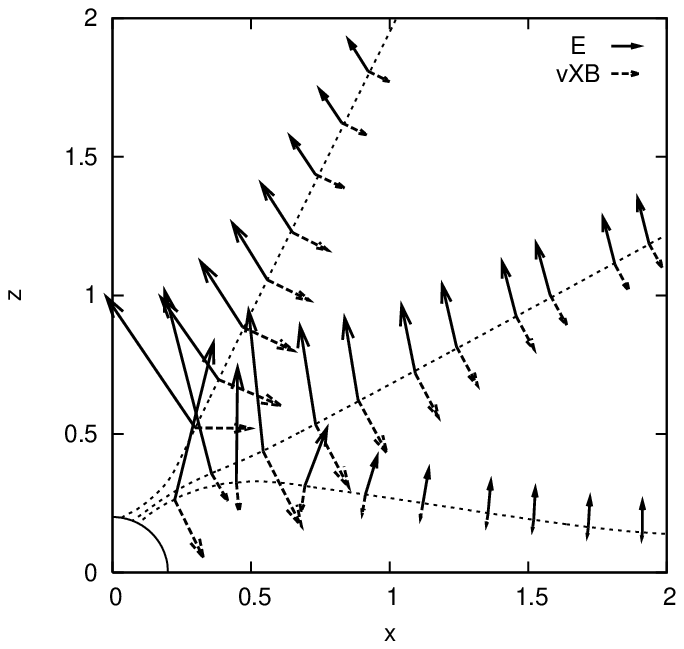}
\caption{ Electromagnetic forces acting on a
positively charged particle on a stream line.
The electric and magnetic forces are shown by  
solid and dotted arrows.
Three dotted curves denote the stream lines.
\label{fig7}
}
\end{minipage}
\end{figure}

%%%
   We discuss a consequence of  
non-ideal MHD field $\vec{B} \cdot \vec{E} \ne 0 $. 
Figure \ref{fig6} shows contour of the Lorentz factor of positively 
charged particles normalized by initial one $\gamma _{0}$.
At the injection boundary, $\gamma _{+}$ is fixed as 
$\gamma _{+}=\gamma _{0}=10^{2}$ for all polar cap angle, 
but there is a gradual increase toward the outer 
radius. The increase is remarkable at low $\theta $, but 
$\gamma _{+}$ is almost constant 
for the flow along the equator. The increase of $\gamma _{+}$ is
determined by the Bernoulli integral as
$\Delta \gamma _{+}=-q\Delta \Phi /(mc^{2})$, 
since $ 0=\Delta K_{+}= \Delta \gamma _{+}+q\Delta \Phi /(mc^{2})$ 
along each flow line. 
Thus, large acceleration of positively charged particles toward 
the polar region can be understood,
since available potential difference $ - \Delta \Phi $ is large
as inferred from Fig.~\ref{fig5}. 
The change of the Lorentz
factor $\gamma _{-}$ of negatively charged particles is 
opposite in the sign, and is given by 
$\Delta \gamma _{-}=q\Delta \Phi /(mc^{2})$. 
They are therefore decelerated toward the polar region.

%%%%% FIG6 %%%%%%%%%%%%%%%%%%%%%%%%%%%%%%%%%%%%
%%%%%%%%%%%%%%%%%%%%%%%%%%%%%%%%%%%%%%%%%%%%%%%

%%%
   Figure \ref{fig7} demonstrates electromagnetic forces 
acting on a positively charged particle on the stream lines.
Two vectors $q\vec{E}$ and $q(\vec{v}\times \vec{B})/c$ 
in the meridian plan are shown by arrows. The sum of these 
two forces causes a net acceleration. It is clear that
the vector $\vec{E}$ is not perpendicular to the flow lines, 
and hence accelerates outwardly.
The acceleration mechanism works better for the flow toward the 
polar region. This is another explanation for the increase of 
the Lorentz factor $\gamma_{+}$ as shown in Fig.~\ref{fig6}.  
This electric effect is opposite for negatively
charged particles, which should be decelerated.

\begin{figure}
%%%%%%%%%%%%%%%%%%
\begin{minipage}{0.45\linewidth}
\includegraphics[scale=1.0]{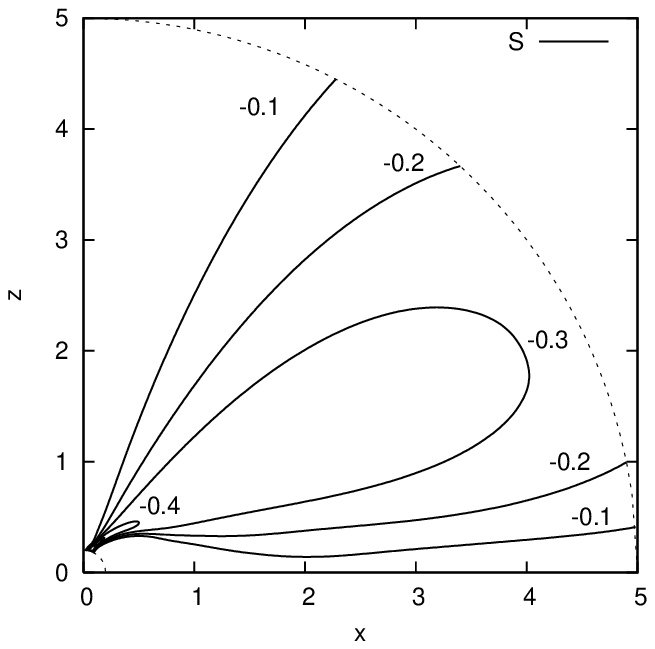}
\caption{  Contour of current stream function  $S$ 
in intervals of $ 0.1 \mu /r_{L} ^2 $.
\label{fig8}
}
\end{minipage}
\hspace{8mm} 
%%%%% FIG9 %%%%%%%%%%%%%%%%%%%%%%%%%%%%%%%%%%%%
\begin{minipage}{0.45\linewidth}
\includegraphics[scale=1.0]{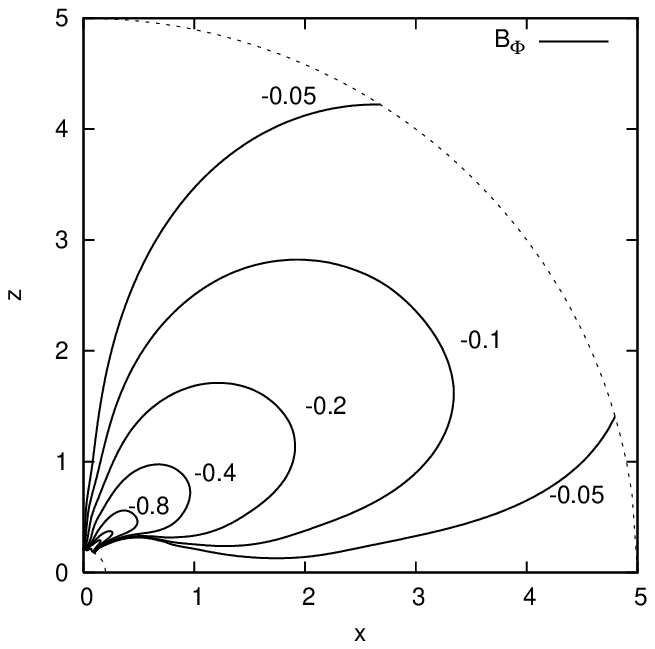}
\caption{ Contour of toroidal magnetic field $B_{\phi}$. 
Contour levels are outwardly for 
$-B_{\phi} r_{L} ^3/ \mu =3.2,1.6,0.8,0.4,0.2,0.1,0.05$.
\label{fig9}
}
\end{minipage}
\end{figure}

%%%
  In Fig.~\ref{fig8}, 
we show the current function $S$. The poloidal current flows along
a curve with a constant value of $S$. 
Figure \ref{fig8} demonstrates a return current. 
That is, two distinct positions at $r_0$ are connected by a curve, 
say, the curve with $S=-0.3 \mu /r_{L}^{2}$.
Such a global return current is generally produced due to 
the inertial term in our model.
The integral $J_{\pm}$ in eq.(\ref{const.Ang}) 
is replaced by $G$ in the limit of $m=0$.
The stream function $F_{\pm}$ is constant on 
a constant magnetic surface $G$, and 
the current function $S= 4 \pi q(F_{+}-F_{-})/c$ is
also constant.
The global structure of $S$ should be the same as that of $G$. 
Therefore, no loop of $S$ is allowed in the limit of $m=0$, 
since $G$ is open field outside corotation region.  
In the context of generalized Ohm's law, the inertial term
is a kind of resistivity. This term causes the dissipation of 
global current, and return current is produced in our model.

%%%
   We consider the effect of the current decay on 
the toroidal magnetic field, which is given by $B_{\phi} = S/R$. 
Figure \ref{fig9} shows the global structure.
The function is zero at the polar axis, on the last open field line
of $G$, and has a maximum at $\theta  \approx \theta_{0}/\sqrt{2}$
of the polar cap region. The function decreases outwardly.
Ratio to the poloidal component is important since
the magnetic field strength both of 
poloidal and toroidal components decreases with radius. 
The ratio is small, 
$|B_{\phi}/ B_{p}| \le (r_{0}/r_{L})^{3/2}/2 \approx 0.04$
at the inner boundary. 
We numerically estimated and found that
$|B_{\phi}/ B_{p}| \approx 0.5 $ at $(r_{L}, \pi/4)$, and
$|B_{\phi}/ B_{p}| \approx 1 $  at  $(4 r_{L}, \pi/4)$. 
Outside the light cylinder, the
poloidal magnetic field is monopole-like as shown in Fig.~\ref{fig2}, 
so that $|B_{p}|\propto r^{-2} $. 
On the other hand, $|B_{\phi}| \propto r^{-1} $ along the
stream lines in the limit of $m=0$.
The slope of $|B_{\phi}| $ slightly becomes steep due
to the inertial term, but is not so steep as $\propto r^{-2} $.
In this way, toroidal component of the 
magnetic field is gradually important with radius, 
although the resistivity is involved in our model.

%%%%% FIG8 %%%%%%%%%%%%%%%%%%%%%%%%%%%%%%%%%%%%
%%%%%%%%%%%%%%%%%%%%%%%%%%%%%%%%%%%%%%%%%%%%%%%

%%%
   The electromagnetic luminosity through a sphere at $r$ is evaluated 
by radial component of the Poynting flux as 
\begin{equation}
L_{\rm{em}} (r) = 2 \int_0 ^{\pi/2} \frac{c}{4 \pi}
 (\vec{E} \times \vec{B} )_{r} 2 \pi r^2 \sin \theta d \theta .
\end{equation}
The luminosity of the plasma flow is a sum of both species as
\begin{equation}
L_{\rm{plasma}}(r) = 2 \int_0 ^{\pi/2} m c^{2}
 ( \gamma_{+} n_{+} v_{+ r} +
  \gamma_{-} n_{-} v_{- r}) 2 \pi r^2 \sin \theta d \theta .
\end{equation}
The energy conversion between two flows is possible 
through the Joule heating $\vec{j} \cdot \vec{E}$, but the total 
$L_{\rm{em}} (r) + L_{\rm{plasma}}(r) $ should be conserved. 
In Table \ref{table2}, numerical results are shown for different radii.
We can check the conservation within a numerical error.
The energy flux by plasma flow is always much larger than the
electromagnetic one, and the conversion is very small in our model. 
The magnitude of the luminosities is almost fixed by the injection 
condition. 
We analytically evaluate these luminosities at $r_{0}$ using
the inner boundary conditions, and find that 
$L_{\rm{em}} = ( \mu \Omega \sin^{2} \theta _{0}/r_{0} )^2 /(3c)$ 
$\approx 0.4 \mu ^2 \Omega ^4/c^3$ and 
$L_{\rm{plasma}} =$
$ 8 \pi m c^2 \gamma_{0} \lambda \sin^{2} \theta _{0}/r_0$
$ =2 \mu ^2 \Omega ^3 \sin^{2}\theta _{0}/(\alpha \eta r_{0})$
$\approx 115 \mu ^2 \Omega ^4/c^3 $, 
where the numerical values $ \alpha =0.2$, $ \eta =0.1$ and 
$\sin^{2}\theta _{0}= 1.15 \Omega /c$ are used.
In order to simulate the Poynting flux dominated case, it is
necessary to increase the parameter $\eta$.

%%%%%%%%%%%%%%%%%%%%%%%%%%%%%%%%%%%%%%%%%%%%%%%%%%%%%%%%%%%%%%%%%%%%%%
% Table 2
%%%%%%%%%%%%%%%%%%%%%%%%%%%%%%%%%%%%%%%%%%%%%%%%%%%%%%%%%%%%%%%%%%%%%%
\begin{table}
\caption{ Luminosity through a sphere with radius $r$. }
\begin{center}
\begin{tabular}{ccc}
\hline\hline
$r/r_{L}$ & $L_{\rm{em}}c^3/(\mu^2 \Omega^4)$ 
& $L_{\rm{plasma}}c^3/(\mu^2 \Omega^4)$ \\ \hline
1.5 & 0.18 & 115.57 \\ 
2.0 & 0.20 & 115.55 \\ 
2.5 & 0.21 & 115.54 \\ 
3.0 & 0.22 & 115.53 \\ 
3.5 & 0.23 & 115.53 \\ 
4.0 & 0.23 & 115.53 \\ 
4.5 & 0.23 & 115.53 \\ \hline\hline
\end{tabular}
\end{center}
\label{table2}
\end{table}
%%%%%%%%%%%%%%%%%%%%%%%%%%%%%%%%%%%%%%%%%%%%%%%%%%%%%%%%%%%%%%%%%%%%%%

%(4)%%%%%%%%%%%%%%%%%%%%%%%%%%%%%%%%%%%%%%%%%%%
  \section{CONCLUSION}
%%%%%%%%%%%%%%%%%%%%%%%%%%%%%%%%%%%%%%%%%%%%%%%
    We have numerically constructed a stationary axisymmetric 
model of magnetosphere with charge separated plasma outflow. 
The stream lines of pair plasma 
are determined by electromagnetic forces and inertial term. 
The massless limit corresponds to the force-free 
and ideal MHD approximations. The global structures of electromagnetic 
fields and plasma flows are calculated by taking into account the 
inertial term. In particular, the non-ideal MHD effects are studied. 
The electrical acceleration or deceleration region 
depending on the charge species appears.
Poloidal current slightly dissipates.
Numerical results depend on a single parameter
$\eta =q\mu /(mc^{2} \gamma _{0} r_{L}^{2})$
as far as $ \gamma _{0} \gg 1$.
The number of our model demonstrated in section 4
is $\eta =0.1$, and is small
when applying to the pulsar magnetosphere.
Typical number $\eta $ is estimated for 
electron-positron pair plasma, 
magnetic field $B_{s}$ at the surface and spin period $P$ as
$\eta =$
$10^{4}(B_{s}/10^{12}\rm{G})(P/1 \rm{s})^{-2}
(\gamma _{0}/10^{2})^{-1}$.
Our present model is only applicable to highly relativistic injection 
($\gamma_{0} \gg 1$) or
weaker magnetic fields ($B_{s}\ll 10^{12}$G).
It will be necessary to scale up many orders of magnitude to 
$\eta \sim 10^{4}$
in order to apply the present method to more realistic cases.
We in fact tried scaling up in our numerical calculations,
but found that it is not straightforward.

%%%
The difficulty and the limitation to smaller value of $\eta $
are closely related with boundary conditions and involved physics,
as explained below.
The Bernoulli integral (\ref{const.Bern}) should satisfy 
a constraint $K_{\pm } \mp q\Psi /(mc^{2}) =\gamma >1 $.
In our model, we specify $K_{\pm } $ 
by the injection condition 
which is fixed at the inner boundary. 
During the numerical iterations, 
the magnitude of the function $\Psi $ 
becomes very large in a certain region, where
the condition 
$K_{\pm } \mp q\Psi /(mc^{2}) >1 $ is no longer satisfied.
It is easily understood that this easily happens for 
large value of  $\eta =q\mu /(mc^{2} \gamma _{0} r_{L}^{2})$,
since the typical scale of  $\Psi $ is $\mu /r_{L}^{2}$.
This gives a certain upper limit to the choice of $\eta $.
The actual estimate of the limit is somewhat complicated, 
since the potential $\Psi $ depends on the choice of boundary
conditions, especially injection model at inner radius.
It is therefore necessary to explore consistent boundary conditions 
or to include some physical process, in order to 
calculate the models with larger $\eta $.
Adjusting mechanism may be required at the boundary
or within the numerical domain.
For example, our present inner boundary is one-way, i.e,
injection only at fixed rate. 
During numerical iterations,
the charge density might numerically blow up elsewhere due to poor
boundary condition.
If the injection rate is able to be adjusted or plasma is absorbed 
through the boundary, the increase may be suppressed.     
However, this is a very difficult back-reaction problem. 
The boundary conditions are normally used 
to determine the inner structures. 
The adjustable boundary conditions should be 
controlled by the interior.
Thus, the boundary conditions and the inner structures
should be determined simultaneously.
Such numerical scheme is not known and should be developed
in future.
Otherwise, extensive study to find out
consistent boundary conditions for large $\eta $
is required.
Our numerical method will be improved by either approach.

%%%
   Pulsar magnetosphere is described in most spatial region 
by the force-free and ideal MHD approximations,
which correspond to the limit of $\eta \gg 1$.
Stationary axisymmetric magnetosphere is constructed so far 
by a solution of the Grad-Shafranov equation with these 
approximations \citep{CKF99,OK03,Gru05}.
Poloidal magnetic field approaches a quasi-spherical wind at infinity.
There is a discontinuity in toroidal magnetic field at the boundary 
of the corotating region, where the current sheet is formed.
\cite{LTR06} have  obtained an alternative solution 
of the same equation, but with a different injection current.
Their solution exhibits a jet along polar axis and a disk on the 
equator. There is no current sheet in their numerical model.
Thus there are at least two models in the strong magnetic field limit, 
$\eta \gg 1$.  
The poloidal magnetic field at infinity is
quasi-spherical and there is no current sheet
in our numerical solution.
It is interesting to examine the model sequence
by increasing the parameter $\eta $.
Some plasma flow should be highly constrained to a thin region,
or the jet-disk system should be formed in the large $\eta $ limit.
It is unclear whether or not many solutions exist, 
depending on physical situations including the plasma state.
In order to address these questions, it
is necessary to construct the magnetosphere with plasma flow 
beyond the force-free and ideal MHD approximations.
We have here presented a possible approach, although 
the improvement is needed.

%%%%%%%%%%%%%%%%%%%%%%%%%%%%%%%%%%%%%%%%%%%%%%%

%%%%%%%%%%%%%%%%%%%%%%%%%%%%%%%%%%%%%%%%%%%%%%%
  \section*{Acknowledgements}
%%%%%%%%%%%%%%%%%%%%%%%%%%%%%%%%%%%%%%%%%%%%
 We would like to thank Shinpei Shibata for valuable discussion. 
This work was supported in part by the Grant-in-Aid for 
Scientific Research (No.16540256 and No.21540271) 
from the Japanese Ministry 
of Education, Culture, Sports, Science and Technology.
%%%%%%%%%%%%%%%%%%%%%%%%%%%%%%%%%%%%%%%%%%%%%%
  
%%%%%%%%%%%%%%%%%%%%%%%%%%%%%%%%%%%%%%%%%

%%%%%%%%%%%%%%%%%%%%%%%%%%%%%%%%%%%%%%%%%%%%%%%
  \appendix
  \section{DIMENSIONLESS FORMS}
%%%%%%%%%%%%%%%%%%%%%%%%%%%%%%%%%%%%%%%%%%%%
 We consider the dimensionless forms of eqs.
(\ref{eqn.Gfn}),(\ref{eqn.Psi}) and (\ref{eqn.trans2}).
The magnetic function $G$ is normalized in terms of 
the magnetic dipole moment $\mu$ and the distance to
light cylinder $ r_{L}$ as $G= \mu  G^{\dagger}/r_{L} $,
where a symbol $^{\dagger}$ denotes a dimensionless quantity.  
The electric potential $\Psi $ is expressed as
$ \Psi =\mu \Omega \Psi ^{\dagger}/(c r_{L} )$
$=\mu \Psi ^{\dagger} / r_{L} ^2 $.
Two integrals (\ref{const.Ang}) and (\ref{const.Bern})
are written as $J_{\pm }= \gamma _{0} c r_{L} J^{\dagger} _{\pm } $
and $K_{\pm }=\gamma _{0} K^{\dagger} _{\pm}$, where
\begin{equation}
J^{\dagger} _{\pm } = \left( \frac{\gamma _{\pm }}{\gamma _{0}} \right)
  \left( \frac{ v_{\pm \phi } }{c} \right)
 \left(  \frac{R}{r_{L}}\right)
 \pm \eta  G^{\dagger} ,
\end{equation}
\begin{equation}
K^{\dagger} _{\pm }=\left( \frac{\gamma _{\pm }}{\gamma _{0}} \right)
\pm \eta \Psi^{\dagger} .
\end{equation}
These values depend on a dimensionless parameter
$\eta =q\mu /(mc^{2}\gamma _{0} r_{L}^{2})$.
As for the stream function $F_{\pm }$,
the normalization constant is 
$\mu \Omega /(q r_{L} )=\mu c/(q r_{L}^2) $ and
$F_{\pm } =\mu \Omega F_{\pm } ^{\dagger} /(q r_{L} )$.
The number density is normalized  from eq.(\ref{dfn.density})
as $n_{\pm } = \mu \Omega n_{\pm } ^{\dagger} /(q c r_{L}^3  )$.
Electric charge and current densities are normalized as
$\rho_{e} = \mu \Omega \rho_{e} ^{\dagger} /(c r_{L}^3  )$
and $\vec{j} = \mu \Omega \vec{j} ^{\dagger} / r_{L}^3 $ .
Using these dimensionless functions,
eqs. (\ref{eqn.Gfn}),(\ref{eqn.Psi}) and  (\ref{eqn.trans2})
can be written as 
\begin{equation}
{\mathcal{D}}^{\dagger} G^{\dagger}
=- 4\pi R^{\dagger} j_{\phi }^{\dagger},
\end{equation}
\begin{equation}
\left( \nabla ^{\dagger } \right) ^{2}  \Psi^{\dagger}  
= -4\pi \left(  \rho _{e}^{\dagger }
-  R^{\dagger } j_{\phi } ^{\dagger}  \right) 
-2 \left(\frac{ 1 }{r ^{\dagger } } \right)^{2}
   \left(  r^{\dagger } \frac{\partial}{\partial r^{\dagger} }
 +\cot \theta \frac{\partial }{\partial \theta } \right) G ^{\dagger} ,
\end{equation}
\begin{equation}
{\mathcal{D}}^{\dagger}  F_{\pm }^{\dagger} 
= \left[ \vec{\nabla} ^{\dagger} 
\ln \left( \frac{n_{\pm }^{\dagger}\gamma_{0} }{\gamma_{\pm }}\right)
+\frac{c^{2}\gamma _{0}  }{\gamma _{\pm }(v_{\pm r}^{2}
+v_{\pm \theta}^{2})}
\left( \vec{\nabla}^{\dagger} K_{\pm }^{\dagger}
-\frac{v_{\pm \phi }}{cR^{\dagger}}\vec{\nabla}^{\dagger}
J_{\pm } ^{\dagger} \right) \right] \cdot
\vec{\nabla}^{\dagger}  F_{\pm } ^{\dagger}
\pm 4 \pi \eta \frac{n_{\pm }^{\dagger} \gamma_{0} }{\gamma _{\pm }}
\left( F_{+} ^{\dagger} - F_{-} ^{\dagger} \right) .
\end{equation}
where $ R^{\dagger}=R/r_{L}$, $ r^{\dagger}=r/r_{L}$ and
the differential operators with the
symbol  $ ^{\dagger}$ are defined by $ r^{\dagger}=r/r_{L}$.
From these expressions, we find that
$ \eta $ is an important parameter.

%
%%%%%%%%%%%%%%%%%%%%%%%%%%%%%%%%%%%%%%%%%
\end{document}